# Macroscopic electro-optical modulation of solution-processed molybdenum disulfide

**Authors**
Songwei Liu[1], Yingyi Wen[1], Jingfang Pei[1], Xiaoyue Fan[2], Yongheng Zhou[3], Yang Liu[1,4], Ling-Kiu Ng[1], Yue Lin[5], Teng Ma[6], Panpan Zhang[7], Xiaolong Chen[3], Gang Wang[2,*], Guohua Hu[1,*]

**Affiliations**
[1]Department of Electronic Engineering, The Chinese University of Hong Kong, Shatin, N. T., Hong Kong S. A. R., China
[2]Centre for Quantum Physics, Key Laboratory of Advanced Optoelectronic Quantum Architecture and Measurement (MOE), School of Physics, Beijing Institute of Technology, Beijing 100081, China
[3]Department of Electrical and Electronic Engineering, Southern University of Science and Technology, Shenzhen, 518055 China
[4]Shun Hing Institute of Advanced Engineering, The Chinese University of Hong Kong, Shatin, N. T., Hong Kong SAR, 999077 China
[5]CAS Key Laboratory of Design and Assembly of Functional Nanostructures, and State Key Laboratory of Structural Chemistry, Fujian Institute of Research on the Structure of Matter, Chinese Academy of Sciences, Fuzhou, Fujian 350002, China
[6]Department of Applied Physics, Hong Kong Polytechnic University, Hung Hom, Kowloon, Hong Kong S. A. R., China
[7]State Key Laboratory of Information Photonics and Optical Communications, Beijing University of Posts and Telecommunications, Beijing 100876, China

*Correspondence to: ghhu@ee.cuhk.edu.hk, gw@bit.edu.cn

**Abstract**
Molybdenum disulfide ($MoS_2$) has drawn great interest for tunable photonics and optoelectronics advancement. Its solution processing, though scalable, results in randomly networked ensembles of discrete nanosheets with compromised properties for tunable device fabrication. Here, we show via density-functional theory calculations that the electronic structure of the individual solution-processed nanosheets can be modulated by external electric fields collectively. Particularly, the nanosheets can form Stark ladders, leading to variations in the underlying optical transition processes and thus, tunable macroscopic optical properties of the ensembles. We experimentally confirm the macroscopic electro-optical modulation employing solution-processed thin-films of $MoS_2$ and ferroelectric P(VDF-TrFE), and prove that the localized polarization fields of P(VDF-TrFE) can modulate the optical properties of $MoS_2$, specifically, the optical absorption and photoluminescence on a macroscopic scale. Given the scalability of solution processing, our results underpin the potential of electro-optical modulation of solution-processed $MoS_2$ for scalable tunable photonics and optoelectronics. As an illustrative example, we successfully demonstrate solution-processed electro-absorption modulators.

**Introduction**

Tunable semiconductor photonics and optoelectronics play a crucial role in sensing, computing, communications, and manufacturing (*1*). Mono- and few-layer molybdenum disulfide ($MoS_2$), characterized by transition electronic structure and dynamic optoelectronic process has garnered significant interest for the advancement of novel photonics and optoelectronics. Specifically, the quantum confinement in $MoS_2$ arising from its two-dimensional configuration and spontaneous translational symmetry breaking enables convenient engineering of its electronic structure by the local fields (*2, 3*). Consequently, $MoS_2$ emerges as ideal for tunable photonics and optoelectronics development, as demonstrated in state-of-the-art advances (*4, 5*). Though promising, the advances rely on the use of pure phase, long-range ordered crystalline $MoS_2$ prepared by methods like mechanical exfoliation and thus, all face an unresolved challenge of scalability (*6*). In contrast, the solution-processed counterpart offers the advantage of production scalability and facilitates the convenient deposition of large-area thin-films (*7-9*). However, compared to the $MoS_2$ prepared using the alternative methods, the solution-processed $MoS_2$ is a randomly networked ensemble of discrete nanosheets (*10, 11*) – while the individual nanosheets retain the material physics and properties of the pure phase, the ensemble exhibits compromised macroscopic properties with distinctly differentiated materials physics. Therefore, to comprehend the physics and properties of the ensemble, it is essential to consider the individual nanosheets as independent systems, while collectively analyzing the ensemble from the perspective of statistical physics (*12*).

To develop tunable photonics and optoelectronics from solution-processed $MoS_2$, it is necessary to collectively modulate the optical properties of the solution-processed $MoS_2$ ensemble on a macroscopic scale. In this pursuit, a comprehensive understanding of the underlying optical processes is crucial. When subjected to an external stimulus, such as incident light, the optical response of the ensemble is primarily determined by two factors: the structural parameters of the ensemble, and the material physics of the individual nanosheets comprising the ensemble. Specifically, assuming the ensemble consists of *N* nanosheets, each of the individual nanosheets can be characterized by *M* specific parameters, and the optical response of the ensemble can be effectively summarized and described by Eq. (1):

$$I_{ensemble} = \sum_{i}^{N} I_i = \sum_{i}^{N} \left[ (\prod_{j}^{M} \sigma_{ij}) \cdot \varphi_i(F; \mathbf{s}_i) \right], \qquad Eq.(1)$$

where $\sigma_{ij}$ is the element of the ensemble structural parameter matrix $\hat{\boldsymbol{\sigma}} = (\boldsymbol{\sigma}_1, \boldsymbol{\sigma}_2, ..., \boldsymbol{\sigma}_N)^{\mathrm{T}} \in \mathrm{P}^{N \times M}$, and $\boldsymbol{\sigma}_i = (\sigma_{i1}, ..., \sigma_{iM}) \in \mathrm{P}^M$ represents the structural parameters of the *i*-th nanosheet within the ensemble, such as the effective area and the tilting angle of the nanosheet. The function $\varphi_i$ describes the optical response of the *i*-th nanosheet to the external stimulus, specifically the incident radiation. In this context, the incident radiation can be described by **s**, encompassing parameters such as wavelength (i.e. the optical response range), intensity, and incident angle of the radiation. Similarly, $\mathbf{s}_i$ represents the incident radiation received by the *i*-th nanosheet. To modulate the optical response of the ensemble, an external electric field *F* is proposed as a promising tool. Indeed, previous studies have demonstrated the effectiveness of electric fields in modulating the electronic structure and optical properties of two-dimensional materials, including (*13-15*). Given that all individual nanosheets in the ensemble retain a two-dimensional configuration, a macroscopic electric field can induce quantum confinement effects, consequently modulating the optical properties of these nanosheets. The minute variations in the optical properties of the nanosheets can accumulate and result in substantial changes in the collective macroscopic optical properties of the ensemble. Building upon this understanding, we propose that by manipulating the optical responses $\varphi$ of the individual MoS$_2$ nanosheets using electric fields, it is feasible to achieve collective electro-optical modulation of the macroscopic optical response $I_{ensemble}$ of the ensemble. Leveraging this modulation approach, as an illustrative example, we demonstrate electro-absorption modulators from solution-processed MoS$_2$, showcasing a promising pathway towards the development of scalable tunable 2D material photonics and optoelectronics.

## Results

**Optical transition processes in solution-processed MoS$_2$:** To investigate the modulation of the macroscopic optical response of the solution-processed MoS$_2$ ensemble using electric fields, it is important to establish a clear understanding of the fundamental light-matter interaction processes involved. In general, the interaction between the incident radiation and semiconductors, including the solution-processed 2D semiconductor nanosheets and their ensemble, involve key processes including reflection, absorption, transmission (optical process without energy conversion), and inelastic scattering of photons and photoluminescence (optical process with energy conversion), as depicted in Fig. 1A. Specifically, for the solution-processed ensemble, these aforementioned light-matter interactions entail the band-to-band optical transition process in the direct bandgap monolayer MoS$_2$ nanosheets (Fig. 1B), and the phonon-assisted optical transition process in the indirect bandgap few-layer MoS$_2$ nanosheets (Fig. 1C).

In our study, we investigate the light-matter interactions in solution-processed MoS$_2$ nanosheets using density-functional theory (DFT) calculations (Supplementary Note 1). Figure S1 illustrates the band-to-band optical transition process in the direct bandgap MoS$_2$ monolayer nanosheets that occurs between the conduction band minimum (CBM) and valence band maximum (VBM) located at $K^{\pm}$. Specifically, Fig. S2 shows that the calculated conduction band edge and valence band edge states are primarily composed of the $d_{z^2}$ states and $d_{x^2-y^2} \pm id_{xy}$ states of the molybdenum (Mo) atoms, respectively, both mixed with the $p_x \mp ip_y$ states of the sulfur (S) atoms.

On the other hand, as illustrated in Fig. S1, our calculations reveal that the phonon-assisted optical transition process in the indirect bandgap few-layer MoS$_2$ nanosheets occurs between the CBM at the halfway point $\Lambda_{min}$ of $\Gamma - K$ path and VBM at $\Gamma$. Considering that the reported studies indicate an average thickness of around 4-5 layers for the solution-processed ensemble (*16, 17*), we calculate and present the electronic structure of penta-layer MoS$_2$ as a representative of the ensemble in Fig. 1D. As observed, the VBM and CBM are located at $\Gamma$ and $\Lambda_{min}$, respectively. To gain a more explicit understanding of the phonon-assisted optical transition processes in the averaged penta-layer MoS$_2$ nanosheets, we present in Fig. 1E-G the calculated electronic structures from the Mo and S atoms as well as the density of states (DOS). It can be observed that the CBM at $\Lambda_{min}$ is primarily composed of Mo $d_{x^2-y^2} \pm id_{xy}$ states mixed with S $p_x \mp ip_y$ orbitals, while the VBM at $\Gamma$ is mainly contributed by Mo $d_{z^2}$ states and S $p_z$ orbitals. It is worth noting that our above calculation results align with previous theoretical and experimental studies conducted on the pure phase crystalline MoS$_2$ produced by the other means (*18*).

**Formation of the Stark ladders in few-layer MoS$_2$:** When an electric field $F$ is applied to the solution-processed MoS$_2$ nanosheets and their ensemble, the electronic structures of the individual nanosheets can easily get split, leading to a reduction in the bandgap. This phenomenon is known as the quantum confined Stark effect (QCSE) (*15, 19*). The electronic structures and their evolution due to the QCSE exhibit a dependence on the thickness of the nanosheets. In the case of few-layer MoS$_2$ in the hexagonal (2H) phase, the inherited intra-layer spacing from the bulk allows the formation of intra-layer dipole moments, as illustrated in Fig. 2A. Once these intra-layer dipoles are formed, the redistribution of the charges in response to the external electric field may become more significant, rendering more intense QCSE, as demonstrated in our calculations (Fig. 2B). Furthur, our calculations show that this enhancement in QCSE can increase with the thickness of the MoS$_2$ nanosheets (Fig. 2C). On the contrary, the mono-layer nanosheets lack an intra-layer spacing, preventing the formation of effective intra-layer dipoles and resulting in a less prominent QCSE (Fig. 2C).

Therefore, based on the above understandings, while the macroscopic optical properties of the ensemble can be governed by the mono-layer nanosheets due to their direct band-to-band optical transition, the collective changes in the optical response induced by the electric field can be primarily attributed to the few-layer nanosheets within the ensemble, as described in Eq. (2):

$$I_{ensemble}(F) = I_{mono-layer} + I_{few-layer}(F)\,. \qquad Eq.(2)$$

Focusing on the few-layer nanosheets, the application of an electric field can induce a progressive band splitting (Fig. 2D), where each of the atomic layers occupies the states at different energy levels (Fig. S3). This progressive band splitting eventually leads to the formation of step-like band structures, known as *Stark ladders* (Fig. 3A).

**Absorption coefficient variation near the absorption edges:** The emergence of the Stark ladders can significantly alter the optical processes in the solution-processed few-layer MoS$_2$ nanosheets (*20, 21*). Consequently, the application of an external electric field can modify their optical processes, thereby collectively modulating the optical responses of the ensemble. Without considering the excitonic effects, the QCSE can be regarded as an extreme quantization case of the Franz-Keldysh effect in thin-slab structured materials, such as the solution-processed few-layer MoS$_2$ nanosheets (*22*). Therefore, the optical absorption coefficient for photons with energy less than the bandgap $E_g$ can be expressed as Eq. (3) (*23, 24*):

$$\alpha(\omega, F_z) = \frac{1}{2}\alpha(\omega, 0)\exp\left[-\frac{4}{3}\left(\frac{E_g - \hbar\omega}{\hbar\theta_z}\right)^{\frac{3}{2}}\right], \qquad Eq.(3)$$

where $\alpha(\omega, 0)$ is the initial optical absorption coefficient in the absence of an electric field, and $\hbar\theta_z = (e^2 F_z^2 \hbar^2/2\mu_z)^{\frac{1}{3}}$ is a function of the longitudinal electric field $F_z$ (Supplementary Note 4). The progressive band splitting and formation of the Stark ladders in the few-layer nanosheets can result in a red-shift of the optical absorption edges of the ensemble, as schematically illustrated in Fig. 3B. According to Eq. (3), the red-shift will lead to changes in the intensity of the optical absorption and reflection for light with wavelength $\lambda > hc/E_g$ ($h = 2\pi\hbar$ is the Planck constant, and $c$ is the speed of light), and the most significant changes will occur around the initial optical absorption edges.

To experimentally validate our theoretical findings, we design an *in situ* electro-optical platform (Fig. 3C) for characterizing solution-processed MoS$_2$ thin-films. Here, the MoS$_2$ is exfoliated via electrochemical exfoliation following a previous report (*17*). Material characterizations confirm that the exfoliated MoS$_2$ is in the pristine 2H phase (Fig. S4). As shown in Fig. 3C, for the convenience of characterization, the MoS$_2$ thin-films is prepared in the ITO/SiO$_2$/MoS$_2$/P(VDF-TrFE)/Au configuration (Supplementary Note 2). The ferroelectric P(VDF-TrFE) is deposited to provide the localized polarization fields for the Stark modulation of the MoS$_2$ (*25*). SiO$_2$ is evaporated to prevent current flows. Using our *in situ* electro-optical setup, we investigate the field-induced modulation of the macroscopic optical properties of the solution-processed MoS$_2$ (Supplementary Note 3). The characterization results reveal that the absorption coefficient of the MoS$_2$ thin-film varies in response to the bias applied (Fig. 3D). Notably, for the photons with energy less than 2.1 eV, the MoS$_2$ thin-film exhibits an increased absorption. The maximum change in the absorption coefficient occurs at around 1.95 eV, which is between the A excitonic absorption peak (~1.84 eV) and the B excitonic absorption peak (~2.00 eV) of MoS$_2$. Therefore, the experimental results support our theoretical findings that the optical properties of the solution-processed MoS$_2$ ensemble can vary around the absorption edges due to QCSE.

**Electric field induced photoluminescence quenching:** The optical processes discussed so far do not involve energy conversion, such as photon-electron, photon-phonon, or photon-phonon-electron interactions. However, the evolution of the electronic structure of the individual MoS$_2$ nanosheets due to QCSE is likely to impact the energy conversion processes. Here, we investigate the photoluminescence phenomenon as a demonstration. Unlike the optical absorption, the photon emission in MoS$_2$ typically occurs through the relaxation of the excited electrons and holes to the conduction band minimum (CBM) and valence band maximum (VBM), respectively, followed by their recombination and subsequent photon emission (*26, 27*). This is because the timescale for intra-band relaxation in MoS$_2$ is much shorter than that of the inter-band transition (*26*). The excited electron-hole pairs, however, tend to separate driven by the electric fields applied, as schematically illustrated in Fig. 4A and B (*28*). As a result, the photon emission, which is measured as the photoluminescence intensity from our solution-processed MoS$_2$ thin-film, may be decreased by the applied bias.

Theoretically, when considering that the electro-optical processes of the solution-processed MoS$_2$ nanosheets and ensemble are likely dominated by the indirect bandgap few-layer components, the excitonic recombination rate of the solution-processed ensemble can be expressed by Eq. (4) (*24*):

$$\gamma_{c \to v} = \frac{2\pi}{\hbar} \cdot \left(\frac{eA_0}{m}\right)^2 \sum_{l,\mu} \left|M_{cv,\parallel}^{l,\mu} \cdot S_{cv,\perp}\right|^2 \left[\sum_{\eta} \left|\phi_\eta(0)\right|^2 \delta\left(\left(E_g + E_\eta\right) - \left(\hbar\omega + \hbar q_\mu\right)\right)\right], \quad Eq.(4)$$

where $A_0$ is the intensity amplitude of the emitted radiation, $\phi_\eta(0)$ is the wavefunction of the exciton with zero relative motion, $E_\eta$ is the energy of the exciton, $\hbar q_\mu$ the phonon energy of mode $\mu$, and $M_{cv,\parallel}^{l,\mu}$ is the transition dipole moment of the exciton-photon coupling. As the longitudinal translational symmetry is broken, the interaction is limited to the two-dimensional atomic plane ($xOy$ plane), while the interaction in the longitudinal direction ($z$-direction) $S_{cv,\perp}$ is decided by the overlap integral of the highest occupied molecular orbitals (HOMO) and the lowest unoccupied molecular orbitals (LUMO), as described by Eq. (5):

$$S_{cv,\perp} = \int_{L_z} \phi_c^* \phi_v dz = \int_{L_z} \phi_c \phi_v dz, \qquad Eq.(5)$$

where $\phi_c$ and $\phi_v$ is the wavefunction of LUMO and HOMO, respectively, and $L_z$ is the thickness of the nanosheets (Supplementary Note 5). As revealed by our calculations (Fig. 4C), in the presence of an external electric field and under constant conditions, the HOMO and LUMO in the solution-processed MoS$_2$ nanosheets will undergo a spatial separation, so as the overlap integral $S_{cv,\perp}$, leading to the decrease of the excitonic recombination rate and thus, a photoluminescence quenching. Note that the photoluminescence quenching induced by an external electric field has been reported in other nanostructured materials (*28, 29*).

To investigate whether the photoluminescence quenching occurs in our solution-processed MoS$_2$ thin-film, we perform *in situ* photoluminescence characterization. As shown in Fig. 4D, the measured photoluminescence intensity decreases as the bias increases. Specifically, as observed, the two characteristic photoluminescence peaks corresponding to the emission of the A and B excitons in MoS$_2$ (at ~1.80 eV and ~1.98 eV, respectively) exhibit significant reductions. It is important to note that the measured full width at half maximum of the photoluminescence peaks (~70 nm) is considerably broader than those reported for the mono- and few-layer MoS$_2$ crystalline samples (*30*). This broadening is attributed to the mixture of nanosheets with various thicknesses in the solution-processed MoS$_2$ ensemble. Nevertheless, the *in situ* electro-photoluminescence characterization confirms that the intensities of A and B excitonic photoluminescence emissions in the solution-processed MoS$_2$ thin film decrease with the applied bias. This observation proves that the electric fields can effectively modulate the collective energy conversion processes in the solution-processed MoS$_2$ nanosheets and ensemble.

**Solution-processed electro-absorption modulators:** An effective macroscopic electro-optical modulation of the solution-processed MoS$_2$ sheds light on tunable optical device development. To demonstrate the feasibility, as an illustrative example, we fabricate electro-absorption modulators (EAMs). Capable of performing active modulation of the optical fields with bias, EAMs are a dispensable technology for datacom and optical computing systems (*31, 32*). However, current EAMs are delicate optical devices demanding high-precision manufacturing of high-crystalline optical materials, such as silicon, lithium niobate, and III-V semiconductors. This contributes to the high cost of the EAMs, a stringent barrier limiting the widespread applications of the EAMs in industry (*33, 34*). Continuous efforts are being made to explore the materials and fabrication methods for scaled-up, more accessible EAMs.

Here, following our aforementioned thin-film development method, we fabricate the EAMs in the ITO/SiO$_2$/MoS$_2$/P(VDF-TrFE)/Au configuration with optimization of the fabrication; Fig. 5A (see also Fig. S5). As observed in Fig. 5B-D, after optimization, the EAMs develop distinctly clear layer-by-layer interfaces. Upon *in situ* electro-optical characterization, we select the reflective mode, as the back electrode of gold is opaque to the visible light. This gives the EAMs a zero transmissivity ($T$), while the transparent front electrode of ITO and insulating layer of SiO$_2$ enable the measurement of the reflection signals from the MoS$_2$ thin-film in the EAMs. Neglecting the minimal optical absorption by ITO and SiO$_2$, the optical reflectivity ($R$) and absorptivity ($A$) of the MoS$_2$ thin-film satisfy $R + A = 1$. Consequently, the change in the absorption of the MoS$_2$ thin-film can be interpreted as the change in the reflection using Eq. (6):

$$\Delta I_{reflected}(V, \lambda) = I_{reflected}(V, \lambda) - I_{reflected}(0, \lambda) = -\Delta I_{absorbed}(V, \lambda). \qquad Eq. (6)$$

where $I_{reflected}$ denotes the measured intensity of the reflected light with a wavelength of $\lambda$, and $\Delta I$ represents the intensity changes in the optical absorption/reflection processes.

As shown in Fig. 5E, the reflectance of a typical MoS$_2$ EAM decreases in response to the applied bias. This is consistent with the predicted rise in the absorption coefficient according to our theory above. More explicitly, as shown in Fig. 5F, the spectrum of the static modulation depth ($\Delta R/R_0$) as calculated from Fig. 5E indicates that the EAM device exhibits a reflection modulation for the incident light with wavelength greater than 575 nm (corresponding to a photon energy around and below the characteristic A and B excitons). Further, as observed, the modulation strength is most pronounced around the absorption edges (near the A and B exciton locations), and it increases with the applied bias, as shown in Fig. 5G. Note that the maximal modulation depth of the EAM device is about 4.6 % at ~ 638 nm, corresponding to the maximum change in the absorption coefficient at 1.95 eV (Fig. 5F). As such, the maximal extinction ratio of the EAM device can be estimated as $(R_{on/off})_{max} = 4.343[\alpha_{50V}(1.95\text{ eV}) - \alpha_{0V}(1.95\text{ eV})]L$, where $L$ is the thickness of the $MoS_2$ thin-film (~100 nm). The estimation yields an extinction ratio of ~8.25 dB. The relative static modulation efficiency $[(R_{on/off})_{max}/\Delta V]$ is thus estimated as ~0.165 dB/V. While proving the functionality of the illustrative example of EAMs, the exhibited optical modulation performance is not yet comparable to that reported in state-of-the-art (*35, 36*). Much work is needed to improve the modulation depth, the extinction ratio, and the modulation efficiency via advancements in materials processing, device fabrication, and potentially metasurface designs (*37*). Nevertheless, the EAMs prove the feasibility of macroscopic electro-optical modulation of solution-processed $MoS_2$ for scaled-up tunable photonics and optoelectronics.

**Discussion**
In this work, we have demonstrated macroscopic electro-optical modulation of solution-processed $MoS_2$. Our theoretical studies cooperated with DFT calculations have revealed that the randomly networked ensembles of discrete solution-processed $MoS_2$ nanosheets can undergo macroscopic evolution in the underlying optical properties in external electric fields, originating from the collective electronic structure evolution in the individual solution-processed nanosheets. Through further experimental investigations, we have proved the effectiveness of this electro-optical modulation approach on the optical absorption and photoluminescence of solution-processed $MoS_2$ on a macroscopic scale. Given the resemblance of 2D materials in their electronic structures, we assume that the macroscopic electro-optical modulation approach may not be limited to the solution-processed $MoS_2$, and can be extended to the other solution-processed 2D materials, such as the other transition metal dichalcogenides and phosphorenes. This demands further studies.

To explore the potential of this macroscopic electro-optical modulation approach in developing tunable photonics and optoelectronics, we have demonstrated the fabrication of EAMs from solution-processed $MoS_2$. The prototyped EAMs exhibit a static modulation depth of 4.6 %, an extinction ratio of 8.25 dB, and a static modulation efficiency of 0.165 dB/V. Although further improvements are required to enhance the optical modulation performance, we believe this prototyping effort of the EAMs has proved the feasibility of developing tunable photonics and optoelectronics from the solution-processed $MoS_2$ and the other related 2D materials. Leveraging the scalability of solution processing, our macroscopic electro-optical modulation approach holds significant potential in developing scaled-up tunable photonics and optoelectronics that can find promising applications in information transportation and processing in optical communication and computing systems.

**References**
[1] Reed, G. T., et al. Silicon optical modulators. *Nat. Photonics* **4**, 518-526 (2010).
[2] Novoselov, K. S., et al. 2D materials and van der Waals heterostructures. *Science* **353**, 6298 (2016).
[3] Ahn, C. H., et al. Electrostatic modification of novel materials. *Rev. Mod. Phys.* **78**, 1185 (2006).
[4] Xia, F., et al. Two-dimensional material nanophotonics. *Nat. Photonics* **8**, 899-907 (2014).
[5] Mak, K. F. and Shan, J. Photonics and optoelectronics of 2D semiconductor transition metal dichalcogenides. *Nat. Photonics* **10**, 216-226 (2016).

[6] Wang, S. Y., et al. Two-dimensional devices and integration towards the silicon lines. *Nat. Mater.* **21**, 1225-1239 (2022).
[7] Zou, T. Y. and Noh, Y. Y. Solution-processed 2D transition metal dichalcogenides: materials to CMOS electronics. *Acc. Mater. Res.* **4**, 548-559 (2023).
[8] Pinilla, S., et al. Two-dimensional material inks. *Nat. Rev. Mater.* **7**, 717-735 (2022).
[9] Conti, S., et al. Printed transistors made of 2D material-based inks. *Nat. Rev. Mater.* **8**, 651-667 (2023).
[10] Kelly, A. C., et al. The electrical conductivity of solution-processed nanosheet networks. *Nat. Rev. Mater.* **7**, 217-234 (2022).
[11] Nasilowski, M., et al. Two-dimensional colloidal nanocrystals. *Chem. Rev.* **116**, 10934-10982 (2016).
[12] Gerosa, M. and Bottani, C. E. Multiple light scattering and near-field effects in a fractal treelike ensemble of dielectric nanoparticles. *Phys. Rev. B* **87**, 195312 (2013).
[13] Klein, J., et al. Stark effect spectroscopy of mono- and few-layer $MoS_2$. *Nano Lett.* **16**, 1554-1559 (2016).
[14] Liu, Y. P., et al. Gate-tunable giant Stark effect in few-layer black phosphorus. *Nano Lett.* **17**, 1970-1977 (2017).
[15] Leisgang, N., et al. Giant Stark splitting of an exciton in bilayer $MoS_2$. *Nat. Nanotechnol.* **15**, 901-907 (2020).
[16] Hu, G. H., et al. A general ink formulation of 2D crystals for wafer-scale inkjet printing. *Sci. Adv.* **6**, 33 (2020).
[17] Lin, Z. Y., et al. Solution-processable 2D semiconductors for high-performance large-area electronics. *Nature* **562**, 254-258 (2018).
[18] Wang, G., et al. Excitons in atomically thin transition metal dichalcogenides. *Rev. Mod. Phys.* **90**, 021001 (2018).
[19] Ramasubramaniam, A., Naveh, D. and Towe, E. Tunable band gaps in bilayer transition-metal dichalcogenides. *Phys. Rev. B* **84**, 205325 (2011).
[20] Berghoff, D., et al. Low-field onset of Wannier-Stark localization in a polycrystalline hybrid organic inorganic perovskite. *Nat. Commun.* **12**, 5719 (2021).
[21] Mendez, E. E., Agullorueda, F. and Hong, J. M. Stark localization in GaAs-GaAlAs superlattices under an electric field. *Phys. Rev. Lett.* **60**, 2426 (1988).
[22] Miller, D. A. B., Chemla, D. S. and Schmittrink, S. Relation between electroabsorption in bulk semiconductors and in quantum wells: The quantum-confined Franz-Keldysh effect. *Phys. Rev. B* **33**, 6976 (1986).
[23] Hader, J., Linder, N. and Dohler, G. H. **k·p** theory of the Franz-Keldysh effect. *Phys. Rev. B* **55**, 6960 (1997).
[24] Hamaguchi, C., *Basic semiconductor physics (3rd edition)*. (Springer, 2017).
[25] Wu, G. J., et al. Programmable transition metal dichalcogenide homojunctions controlled by nonvolatile ferroelectric domains. *Nat. Electron.* **3**, 43-50 (2020).
[26] Shi, H. Y., et al. Exciton dynamics in suspended monolayer and few-layer $MoS_2$ 2D crystals. *ACS Nano* **7**, 1072-1080 (2013).
[27] Wang, R., et al. Ultrafast and spatially resolved studies of charge carriers in atomically thin molybdenum disulfide. *Phys. Rev. B* **86**, 045406 (2012).
[28] Horikoshi, Y., Fischer, A. and Ploog, K. Photoluminescence quenching in reverse-biased $Al_xGa_{1-x}As$/GaAs quantum-well heterostructures due to carrier tunneling. *Phys. Rev. B* **31**, 7859 (1985).

[29] Mendez, E. E., et al. Effect of an electric-field on the luminescence of GaAs quantum wells. *Phys. Rev. B* **26**, 7101 (1982).
[30] Mak, K. F., et al. Atomically thin MoS$_2$: A new direct-gap semiconductor. *Phys. Rev. Lett.* **105**, 136805 (2010).
[31] Romagnoli, M., et al. Graphene-based integrated photonics for next-generation datacom and telecom. *Nat. Rev. Mater.* **3**, 392-414 (2018).
[32] Sun, Z. P., Martinez, A. and Wang, F. Optical modulators with 2D layered materials. *Nat. Photonics* **10**, 227-238 (2016).
[33] Sun, D., et al. Microstructure and domain engineering of lithium niobate crystal films for integrated photonic applications. *Light Sci. Appl.* **9**, 197 (2020).
[34] Liu, K., et al. Review and perspective on ultrafast wavelength-size electro-optic modulators. *Laser Photonics Rev.* **9**, 172-194 (2015).
[35] Iannone, P., et al. PAM-4 transmission up to 160 Gb/s with surface-normal electro-absorption modulators. *Opt. Lett.* **45**, 4484-4487 (2020).
[36] Agarwal, H., et al. 2D-3D integration of hexagonal boron nitride and a high-κ dielectric for ultrafast graphene-based electro-absorption modulators. *Nat. Commun.* **12**, 1070 (2021).
[37] Gu, T., et al. Reconfigurable metasurfaces towards commercial success. *Nat. Photonics* **17**, 48-58 (2023).

**Acknowledgements**
**General**: We thank the Shenzhen Cloud Computing Center, National Supercomputing Center and High-Performance Computing Platform of BUPT for providing the high-performance computing clusters services, and Miss Xiaolin Liu for the discussions on the *ab initio* calculations.

**Funding:** GHH acknowledges support from CUHK (4055115), RGC (24200521) and NSFC (62304196), Yang Liu from SHIAE (RNE-p3-21), JFP and YYW from RGC (24200521), Yue Lin from NSFC (52273029), Fujian Science & Technology Innovation Laboratory for Optoelectronic Information of China (2021ZZ119), Pilot Project of Fujian Province (2022H0037) and Natural Science Foundation of Fujian Province for Distinguished Young Scholars (2023J06045), TM from PolyU (P0042991), PPZ from BUPT (Startup Funding), XLC from NSFC (62275117) and Shenzhen Basic Research Program (20220815162316001), and GW from NSFC (12074033).

**Author contributions:** SWL, GW, GHH designed the experiments. SWL, XYF, YHZ, JFP, YYW, Liu Yang, LKN performed the experiments. SWL, Yue Lin, TM, PPZ, XLC, GW, GHH analysed the data. SWL, GHH prepared the figures. SWL, GHH wrote the manuscript. All authors discussed the results from the experiments and commented on the manuscript.

**Competing interests:** The authors declare no competing financial interests.

**Data and materials availability:** The data that support the findings of this study are available from the corresponding authors upon request.

**Figures**

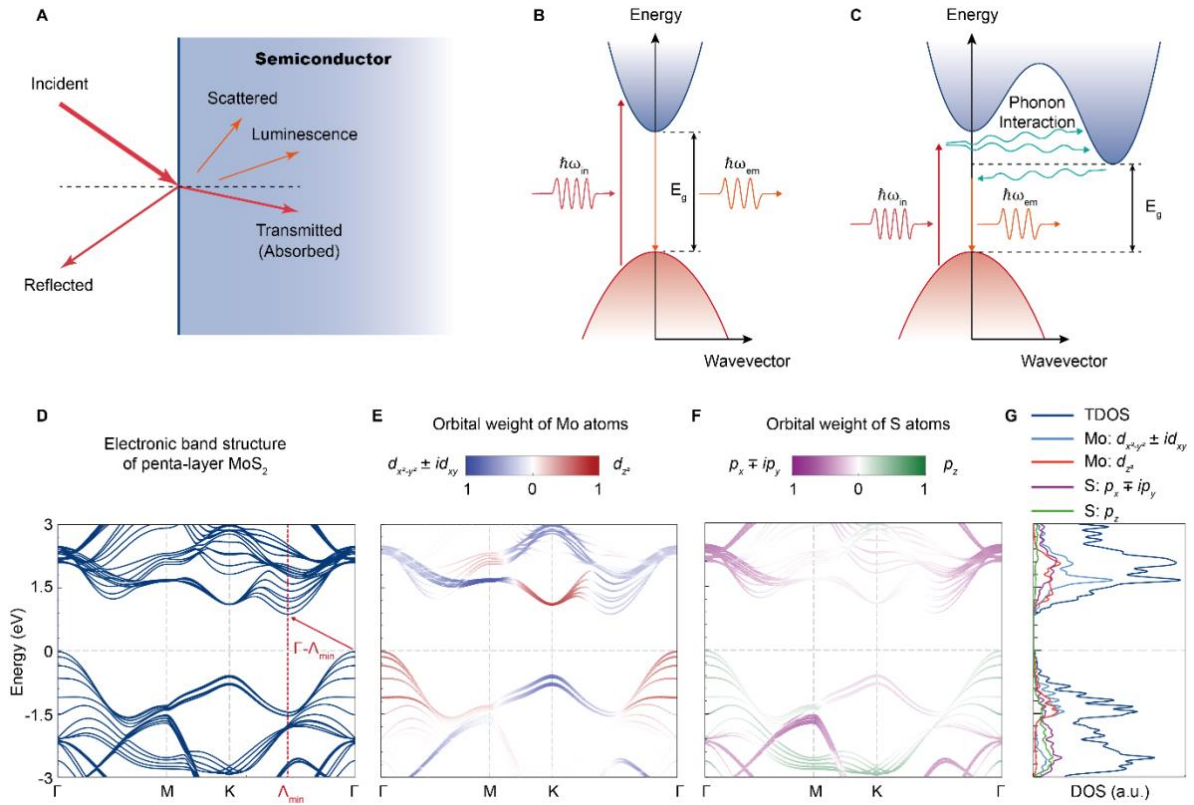

**Figure 1. Optical processes in solution-processed MoS₂.** (A) Schematic linear and nonlinear optical processes in light-semiconductor interactions. (B) Schematic band-to-band optical transition processes in direct bandgap semiconductors, and (C) schematic phonon-assisted optical transition process in indirect bandgap semiconductors. (D) The calculated electronic band structure for penta-layer MoS$_2$ nanosheets, the orbital analysis on the contributions to the electronic structure from the corresponding (E) Mo atoms and (F) S atoms, and the corresponding (G) density of states (DOS), respectively. The computational results from penta-layer MoS$_2$ nanosheets are exampled to represent the collective electronic structure of the solution-processed MoS$_2$ ensemble. See the calculated electronic structures in MoS$_2$ nanosheets of the other thicknesses in Fig. S1.

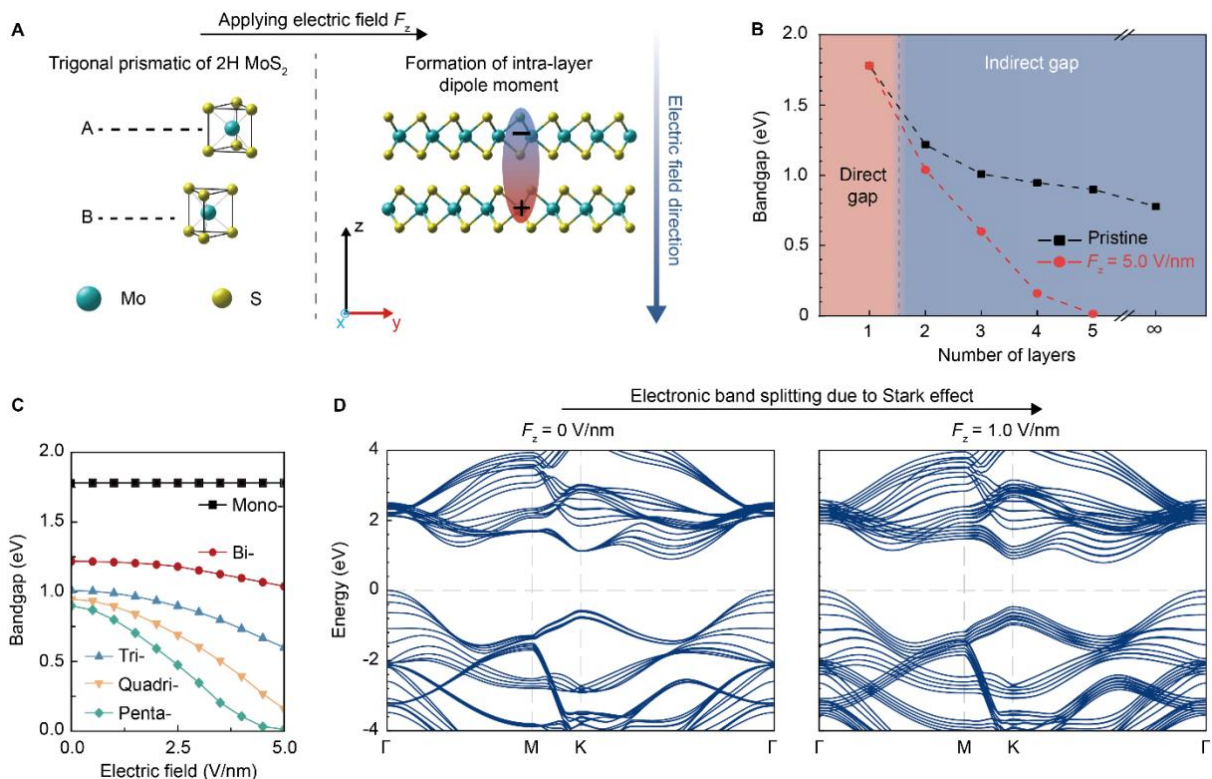

**Figure 2. Electronic structure evolution in solution-processed MoS$_2$ under an external electric field.** (A) Schematic illustration of the atomic structure of hexagonal (2H) MoS$_2$, and the formation of intra-layer dipole moments in response to an external electric field. (B) Calculated bandgap summary of the pristine monolayer, few-layer and bulk MoS$_2$, and the calculated bandgap summary of the MoS$_2$ under an electric field of $F_z = 5.0$ V/nm. (C) Bandgap evolution of monolayer to penta-layer MoS$_2$ nanosheets under an external longitudinal electric field $F_z$. (D) Electronic band splitting of penta-layer MoS$_2$ nanosheets due to the quantum confined Stark effect incurred by the external field.

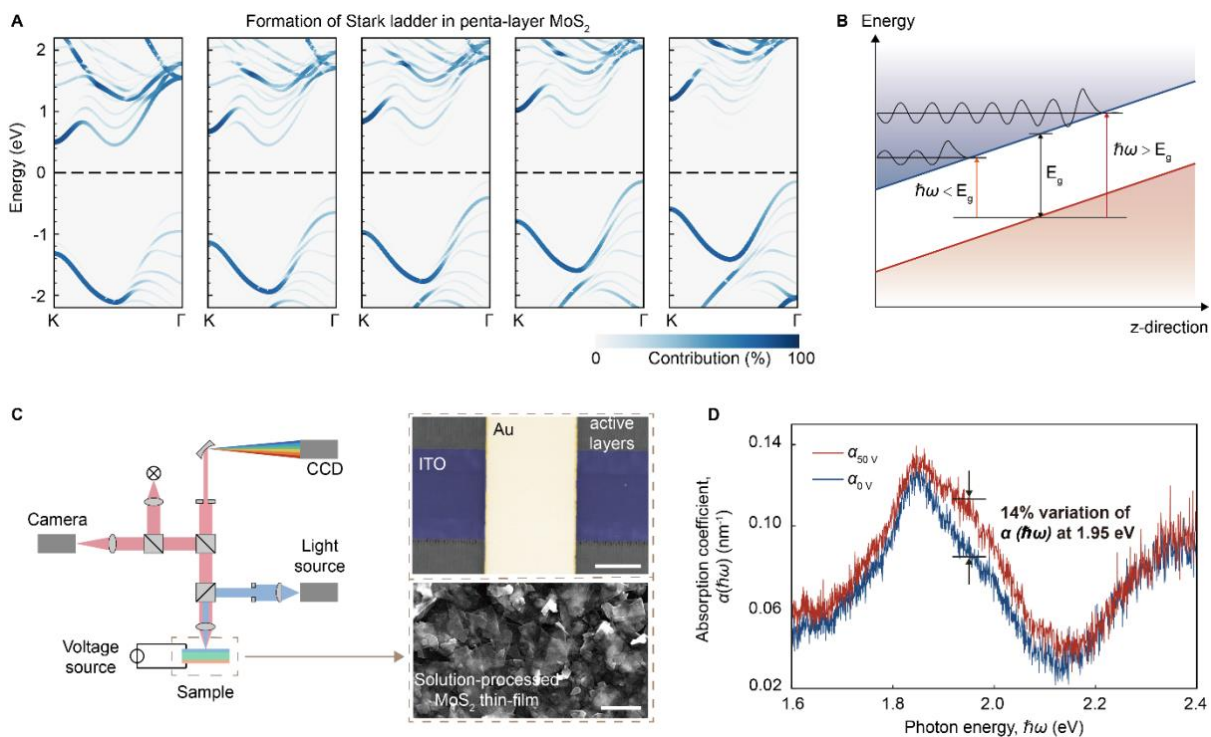

**Figure 3. Electro-optical modulation of solution-processed MoS$_2$.** (A) The formation of the Stark ladder inside the penta-layer MoS$_2$ nanosheets. The computational results from the penta-layer MoS$_2$ nanosheets are exampled to represent the averaged collective electronic structure evolution in the solution-processed MoS$_2$. (B) Schematic Franz-Keldysh effect due to the formation of the *Stark ladders* in response to the electro-optical modulation. (C) Schematic *in suit* electro-optical characterization setup. The *in suit* electro-optical response from the solution-processed MoS$_2$ thin-film sample is measured. The subplots to the right are the top-view optical microscopic image of a typical solution-processed MoS$_2$ thin-film sample for the characterization, and the scanning microscopic image of the deposited MoS$_2$ nanosheets. The active layers are in SiO$_2$/MoS$_2$/P(VDF-TrFE) configurations. Scale bars – (top) 500 µm, (bottom) 500 nm. (D) Optical absorption coefficient spectrum of the solution-processed MoS$_2$ thin-film sample with and without bias.

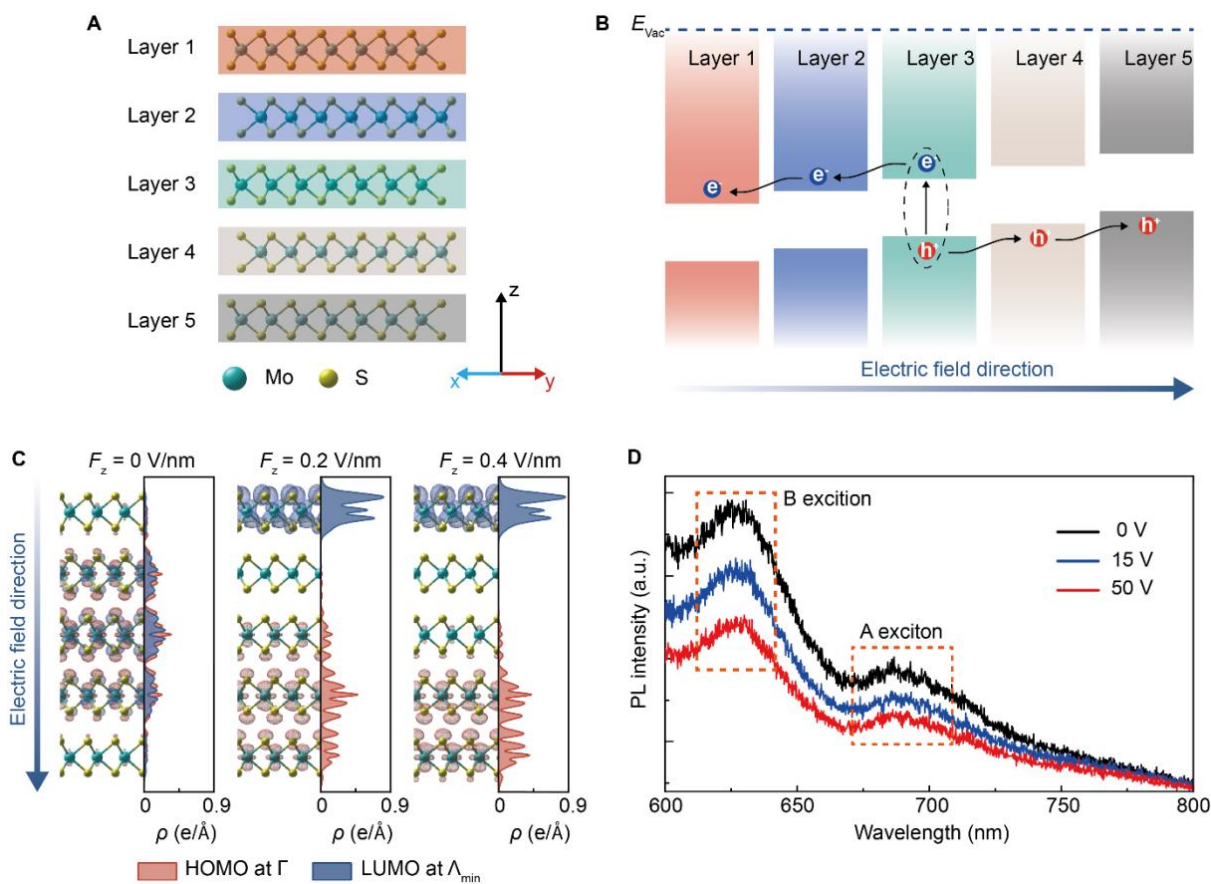

**Figure 4. Field-induced photoluminescence quenching in solution-processed MoS₂.** (A) Structure of penta-layer MoS$_2$ nanosheets. (B) Schematic electron-hole pair separation in a MoS$_2$ nanosheet due to the Stark ladder in an external electric field, showing the separation and propagation processes of an electron-hole pair. (C) The calculated spatial distribution of the HOMO and LUMO in the penta-layer MoS$_2$ nanosheets. Substantial spatial separation takes place in response to the external electric field, showing the redistribution of electrons and holes. (D) *In suit* experimental photoluminescence characterization of the solution-processed MoS$_2$ device sample, showing electro-optical photoluminescence quenching in the solution-processed MoS$_2$ ensemble.

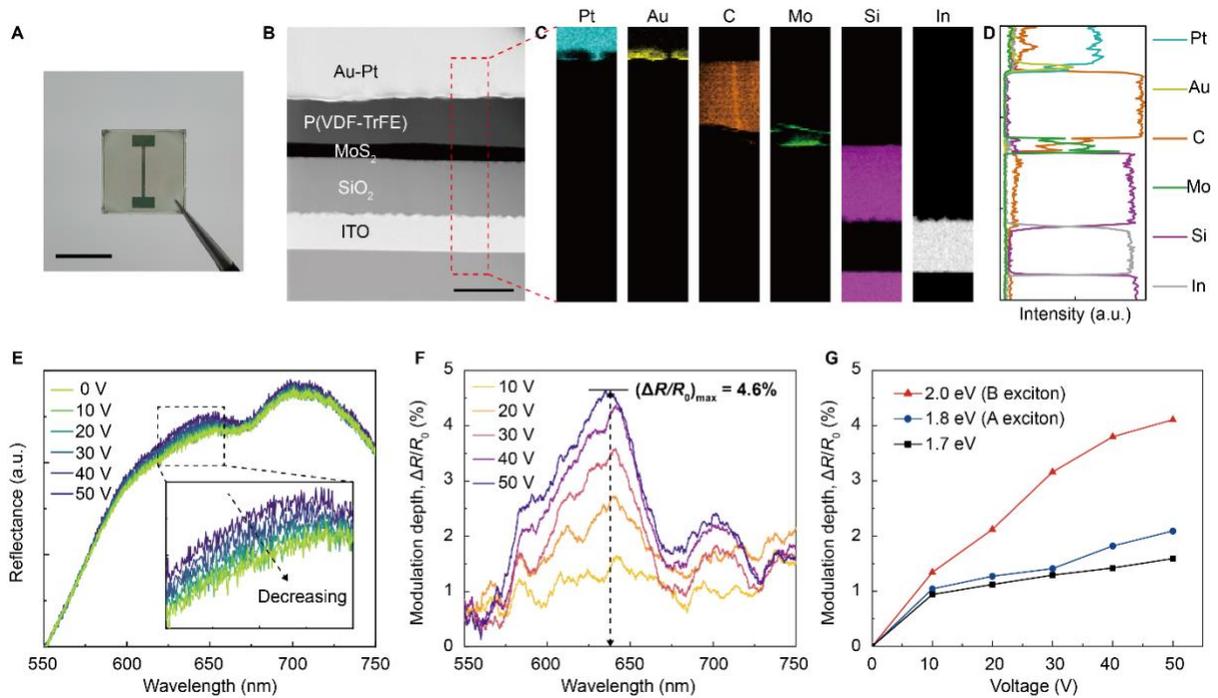

**Figure 5. Solution-processed MoS$_2$ electro-absorption modulators (EAMs).** (A) Picture of a typical EAM device. (B) Cross-sectional scanning transmission electron microscopic image of the device, and the corresponding (C) energy dispersive X-ray area mapping and (D) linear profile. The device is in an ITO/SiO$_2$/MoS$_2$/P(VDF-TrFE)/Au structure. (E) *In suit* electro-reflectance characterization of the EAM device under the different biases, and (F) the corresponding static absorption modulation depth $\Delta R/R_0$ calculated from the *in suit* electro-reflectance characterization. (G) Summary of the modulation depth of the EAM device to the characteristic photon energies, i.e. the A, B excitons and an energy slightly smaller than the absorption edge of the solution-processed MoS$_2$. Scale bars – (A) 10 mm, (B) 500 nm.

Supplementary Materials for
**Macroscopic electro-optical modulation of solution-processed molybdenum disulfide**

**Authors**
Songwei Liu[1], Yingyi Wen[1], Jingfang Pei[1], Xiaoyue Fan[2], Yongheng Zhou[3], Yang Liu[1,4], Ling-Kiu Ng[1], Yue Lin[5], Teng Ma[6], Panpan Zhang[7], Xiaolong Chen[3], Gang Wang[2,*], Guohua Hu[1,*]

**Affiliations**
[1]Department of Electronic Engineering, The Chinese University of Hong Kong, Shatin, N. T., Hong Kong S. A. R., China
[2]Centre for Quantum Physics, Key Laboratory of Advanced Optoelectronic Quantum Architecture and Measurement (MOE), School of Physics, Beijing Institute of Technology, Beijing 100081, China
[3]Department of Electrical and Electronic Engineering, Southern University of Science and Technology, Shenzhen, 518055 China
[4]Shun Hing Institute of Advanced Engineering, The Chinese University of Hong Kong, Shatin, N. T., Hong Kong SAR, 999077 China
[5]CAS Key Laboratory of Design and Assembly of Functional Nanostructures, and State Key Laboratory of Structural Chemistry, Fujian Institute of Research on the Structure of Matter, Chinese Academy of Sciences, Fuzhou, Fujian 350002, China
[6]Department of Applied Physics, Hong Kong Polytechnic University, Hung Hom, Kowloon, Hong Kong S. A. R., China
[7]State Key Laboratory of Information Photonics and Optical Communications, Beijing University of Posts and Telecommunications, Beijing 100876, China

*Correspondence to: ghhu@ee.cuhk.edu.hk, gw@bit.edu.cn

**This file contains:**
Supplementary Notes 1-5
Supplementary Figure S1-S5
Supplementary References

## Supplementary Notes

### Supplementary Note 1: Density-functional theory calculations

In this work, the Vienna *Ab-initio* Simulation Package (*1*) is adopted to perform the density-functional theory calculations. Projector augmented wave pseudopotential (*2*) is used to describe the interaction between the ionic core and valence electrons. The generalized gradient approximation functional in the Perdew-Burke-Ernzerhof scheme (*3*) is used to describe the exchange-correlation (XC) energy between the electrons. The Mo ($4p$, $5s$, $4d$) states and S ($3s$, $3p$) states are treated as valence states. The plane wave cut-off energy is set to be 500 eV. Van der Waals correction is carried out during the crystal structural relaxation using the DFT-D3(BJ) method (*4, 5*). The structure of all the systems is fully optimized until the residual force on the atoms is less than 0.01 eV/Å. For the broken translational symmetry in the solution-processed $MoS_2$ nanosheets, we use a slab geometry with a 22 Å vacuum layer to prevent the interaction between the nanosheets. For the self-consistent iteration, the convergence tolerance between the successive steps is set as $1\times10^{-6}$ eV. Monkhorst-Pack method is used to generate the k-point grid. All the solution-processed $MoS_2$ nanosheets use a $14\times14\times1$, $21\times21\times1$ and $31\times31\times1$ grid for the structural relaxation, self-consistent calculation, and calculation of density of states, respectively. As shown in Fig. 3, the solution-processed $MoS_2$ device is fabricated in a vertical configuration. As such, the local electric fields always have components perpendicular to the solution-processed $MoS_2$ nanosheets except for those minority nanosheets with their atomic planes vertically aligned. Therefore, vertical electric fields are applied by introducing a dipole sheet in the middle of the vacuum layer in this research. The visualization of the atomic structure, molecular orbitals, and charge density are carried out via VESTA (*6*).

### Supplementary Note 2: Preparation of the ITO/SiO$_2$/MoS$_2$/P(VDF-TrFE)/Au device

Solution-processed $MoS_2$ suspension is prepared by electrochemical exfoliation of bulk $MoS_2$ crystals (SPI Supplies), following the method reported in Ref. (*7*). P(VDF-TrFE) (70:30) (Sigma) is dissolved in N,N-dimethylformamide (DMF) (Alfa Aesar) via stirring to form a 2.5 wt.% dispersion. $SiO_2$ is deposited onto the ITO substrate via e-beam evaporation. The solution-processed $MoS_2$ suspension and P(VDF-TrFE) dispersion are then deposited onto the ITO/SiO$_2$ substrate through a layer-by-layer spin-coating process, followed by a heat treatment to remove the residual solvents. Opaque back-contact Au electrode is deposited through e-beam evaporation on the top to complete the device fabrication.

### Supplementary Note 3: *In situ* electro-optical characterizations

The *in situ* electro-optical characterizations are performed at room temperature using a 532 nm continuous laser excitation with a diffraction-limited excitation beam diameter of ~1 μm. The excitation light is filtered by a 532-nm-long pass filter (Semrock). The photoluminescence signal light is separated by a 300 g/mm grating, and then received by a liquid nitrogen-cooled charge-coupled device. The *in situ* electro-optical characterizations under bias are implemented by Keithley 2636B.

### Supplementary Note 4: The Franz-Keldysh effect for two-dimensional (2D) nanosheets

The Franz-Keldysh effect can alter the dielectric property of materials due to the presence of external electric fields (*8*). From a first-principles perspective, this effect originates from the electronic energy band bending under the influence of the external fields (*9*). In general, 2D semiconducting nanosheets including the solution-processed MoS2 nanosheets can exhibit Stark ladders due to the quantum confined Stark effect (QCSE), allowing band bending for the occurrence of Franz-Keldysh effect. Indeed, this is reported in the previous studies that the QCSE can be considered as an extreme quantization case of the Franz-Keldysh effect in a thin-slab structured material (*10*).

Considering the limitation of an individual 2D semiconducting nanosheet, under an electric field $F_z$ perpendicular to the atomic plane, the dielectric function of the nanosheet in this field can be expressed as (*8, 9*):

$$\kappa_2(\omega, F_z) = \frac{1}{2}\kappa_2(\omega, 0)\exp\left[-\frac{4}{3}\left(\frac{E_g - \hbar\omega}{\hbar\theta_z}\right)^{\frac{3}{2}}\right], \qquad Eq.(1)$$

where $E_g$ is the bandgap of the nanosheet, $\kappa_2(\omega, 0)$ is the dielectric function in the absence of the electric field, and $\hbar\theta_z = (e^2 F_z^2 \hbar^2/2\mu_z)^{\frac{1}{3}}$ is a function of the longitudinal electric field $F_z$. Recall the optical absorption coefficient $\alpha = \omega\kappa_2/n_0 c$ (where $\omega$ is the angular frequency of the incident electromagnetic radiation, $n_0$ is the refractive index, and $c$ is the speed of light), we have (*8, 9*):

$$\alpha(\omega, F_z) = \frac{1}{2}\alpha(\omega, 0)\exp\left[-\frac{4}{3}\left(\frac{E_g - \hbar\omega}{\hbar\theta_z}\right)^{\frac{3}{2}}\right]. \qquad Eq.(2)$$

This means that an electron in the valence band can be allowed to be excited into a conduction band by absorbing a photon with an energy below the bandgap.

**Supplementary Note 5: Photoluminescence (PL) quenching in the external electric field**
In a typical carrier excitation process, the electrons can be excited from a valence band lower than the valence band maximum (VBM) to a conduction band higher than the conduction band minimum (CBM) by absorbing a photon with sufficient energy (*11, 12*). The carrier recombination is different from the excitation. The recombination always happens between the electrons at the CBM and the holes at the VBM. This is because the intraband carrier relaxation timescale is much shorter than that of the interband transition (*12*). Given that the effective contribution for the electro-optical modulation of the solution-processed MoS2 ensemble is primarily from the indirect bandgap few-layer nanosheets, we take electron-phonon interactions into account.

We take the limitation of an individual 2D semiconducting nanosheet into consideration, as we discuss above. Defining $H_e$ and $H_p$ as the Hamiltonian of the electrons and lattice vibrations (phonons), respectively, $H_{e-R}$ is the Hamiltonian for the electron-radiation (photon) interaction, and $H_{e-p}$ is the Hamiltonian for the electron-phonon interaction. The total Hamiltonian of a 2D indirect semiconductor nanosheet can therefore be expressed as (*8*):

$$H = H_0 + H' = H_e + H_p + H_{e-R} + H_{e-p}, \qquad Eq.(3)$$

where $H_0 = H_e + H_p$ is the ground state of the system, and $H' = H_{e-R} + H_{e-p}$ is the perturbation. As such, the transition probability $P_{if}$ from an initial state $|i\rangle$ to a final state $|f\rangle$ through a virtual state $|l\rangle$ mediated by a phonon of mode $\mu$ with energy $\hbar q_\mu$ is (*8*):

$$P_{if} = \frac{2\pi}{\hbar}\left|\langle f|H'|i\rangle + \sum_l \frac{\langle f|H'|l\rangle\langle l|H'|i\rangle}{E_i - E_l}\right|^2 \delta(E_i - E_f). \qquad Eq.(4)$$

To describe the electromagnetic fluctuation of the radiative emission, we can use a plane wave $\mathbf{A}(\mathbf{r},t) = A_0 \cos(\mathbf{pr} - \omega t) \cdot \hat{n}$, where $A_0$ is the amplitude of the vector potential, and $\hat{n}$ is the unit vector parallel to $\mathbf{A}$. The term $\omega$ is the angular frequency of the radiation, and $\mathbf{p}$ is the wave vector of the emitted photon.

In general, for the 2D nanosheets, the electronic wavefunctions are confined to form a 2D planar configuration, where the electronic states within the atomic plane can be described by the 2D Bloch function $\psi_\parallel(\mathbf{r}_\parallel) = u_\parallel(\mathbf{r}_\parallel)\exp(i\mathbf{k}_\parallel \mathbf{r}_\parallel)$ (where $\mathbf{r}_\parallel$ is the position in the $xOy$ plane, and $u_\parallel(\mathbf{r}_\parallel)$ is a periodic function). As for the perpendicular direction (i.e. the z-direction), we can use $\phi(z)$ to describe the electronic state distribution. Therefore, the electronic wave function of 2D nanosheets can be expressed as

$$\psi(\mathbf{r}) = \psi_\parallel(\mathbf{r}_\parallel) \cdot \phi(z) = u_\parallel(\mathbf{r}_\parallel)\exp(i\mathbf{k}_\parallel \mathbf{r}_\parallel)\phi(z) \ . \qquad Eq.\,(5)$$

Since the excitonic PL dominates in MoS$_2$ (*13*), taking the excitonic effects into account and by the Fermi's golden rule, we have the exciton recombination rate as (*8*):

$$\gamma_{c\to v} = \frac{2\pi}{\hbar} \cdot \left(\frac{eA_0}{m}\right)^2 \sum_{l,\mu}\left|M_{cv,\parallel}^{l,\mu} \cdot S_{cv,\perp}\right|^2 \left[\sum_\eta |\phi_\eta(0)|^2 \delta\left((E_g + E_\eta) - (\hbar\omega + \hbar q_\mu)\right)\right], \quad Eq.\,(6)$$

where $\phi_\eta(0)$ is the wavefunction of the exciton with zero relative motion, and $E_\eta$ is the energy of the exciton. $S_{cv,\perp}$ is decided by the overlap integral of the highest occupied molecular orbitals (HOMO) and the lowest unoccupied molecular orbitals (LUMO):

$$S_{cv,\perp} = \int_{L_z} \phi_c^* \phi_v dz = \int_{L_z} \phi_c \phi_v dz \ , \qquad Eq.\,(7)$$

where $\phi_c$ and $\phi_v$ is the wavefunction of LUMO and HOMO, respectively, and $L_z$ is the thickness of the nanosheet. As the LUMO and HOMO of the nanosheets separates, $S_{cv,\perp}$ will drop and lead to PL quenching. This electric field induced phenomena have been reported for the other nanostructured material systems in the previous studies (*14, 15*).

**Supplementary Figures**

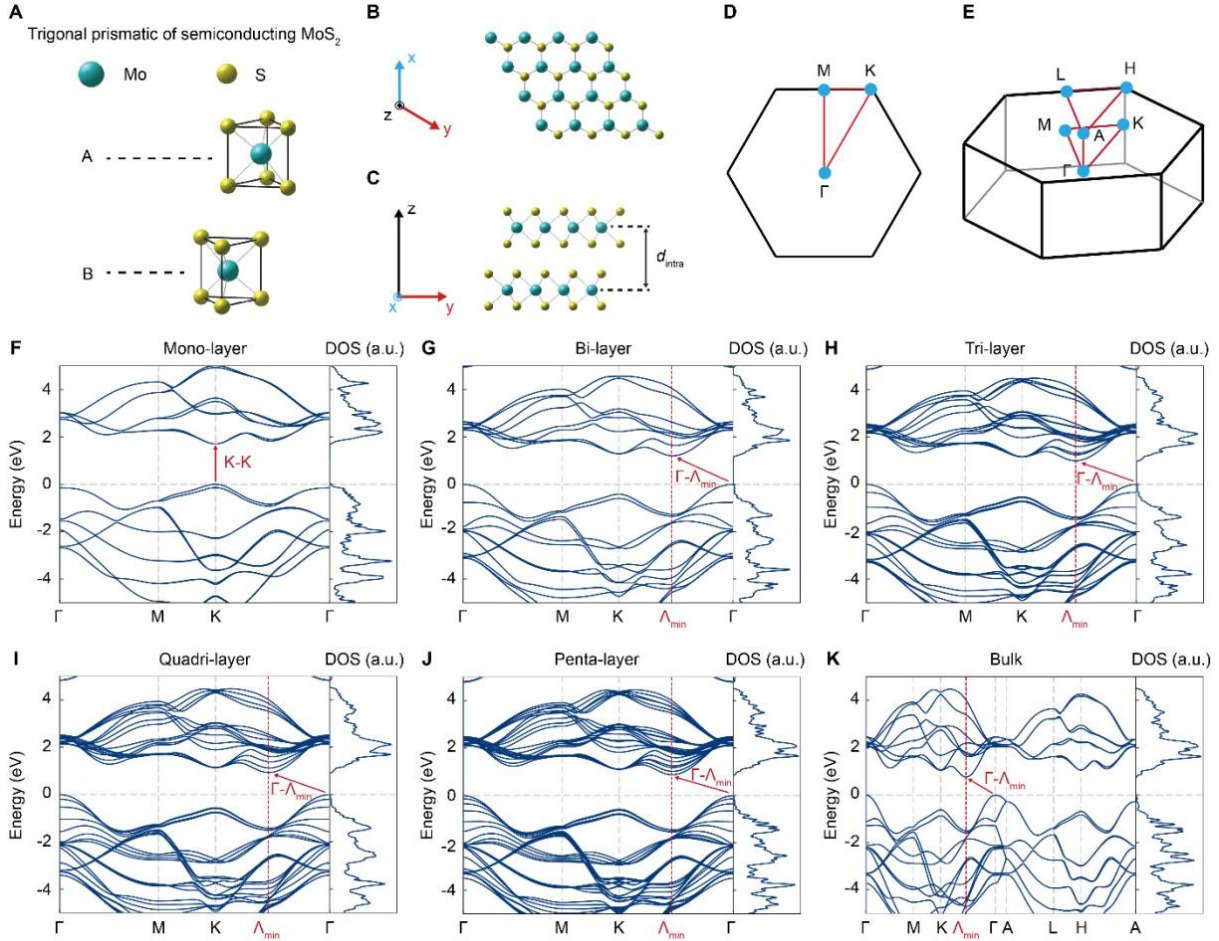

**Figure S1. Atomic and electronic structures of monolayer, few-layer and bulk MoS$_2$ in 2H semiconducting phase.** (A) Atomic structure and the (B) top-view and (C) side-view of semiconducting MoS$_2$ in 2H phase. (D) Two-dimensional (2D) and (E) three-dimensional (3D) Brillouin zones of 2D and 3D hexagonal lattices, respectively. The electronic structures of MoS$_2$ in (F) monolayer, (G-J) few-layer, and (K) bulk configurations.

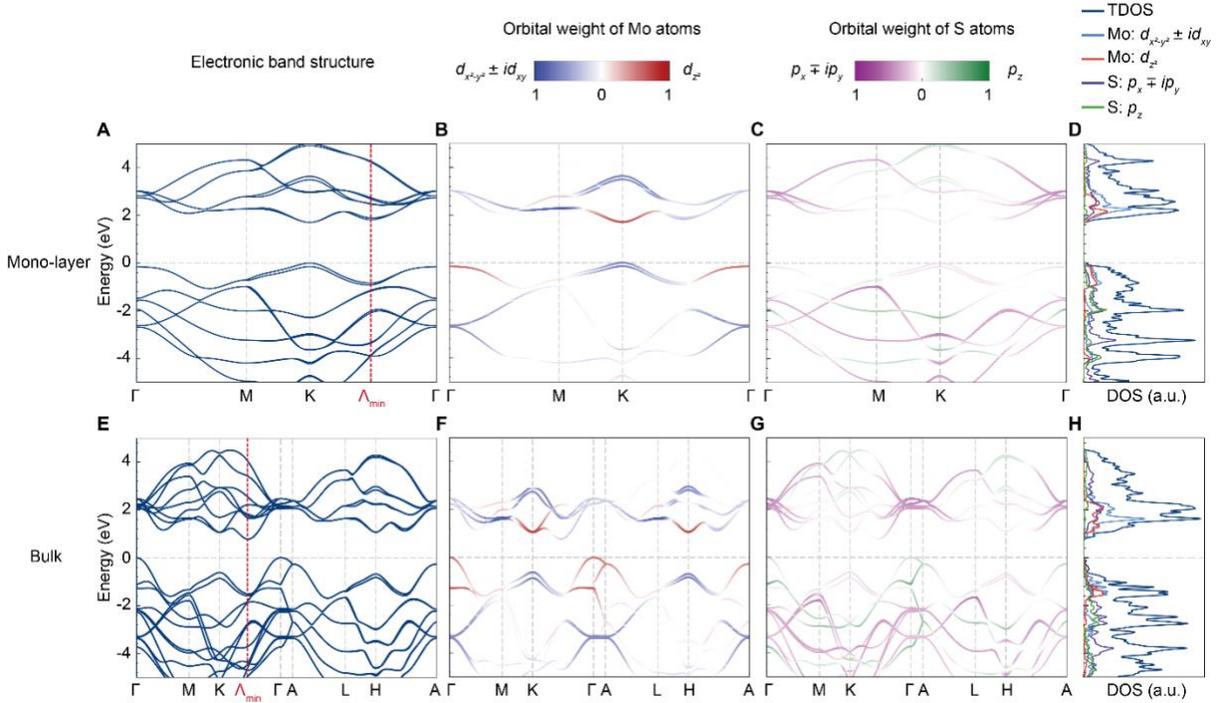

**Figure S2. Electronic structure analysis of monolayer and bulk MoS₂.** (A) Electronic band structure of monolayer MoS$_2$, and the orbital analysis on the contribution to the electronic structure of the (B) Mo atoms and (C) S atoms, respectively. (D) The density of states (DOS) of monolayer MoS$_2$. (E) Electronic band structure of bulk MoS$_2$, and the orbital analysis on the contribution to the electronic structure of the (F) Mo atoms and (G) S atoms, respectively. (H) The density of states (DOS) of bulk MoS$_2$.

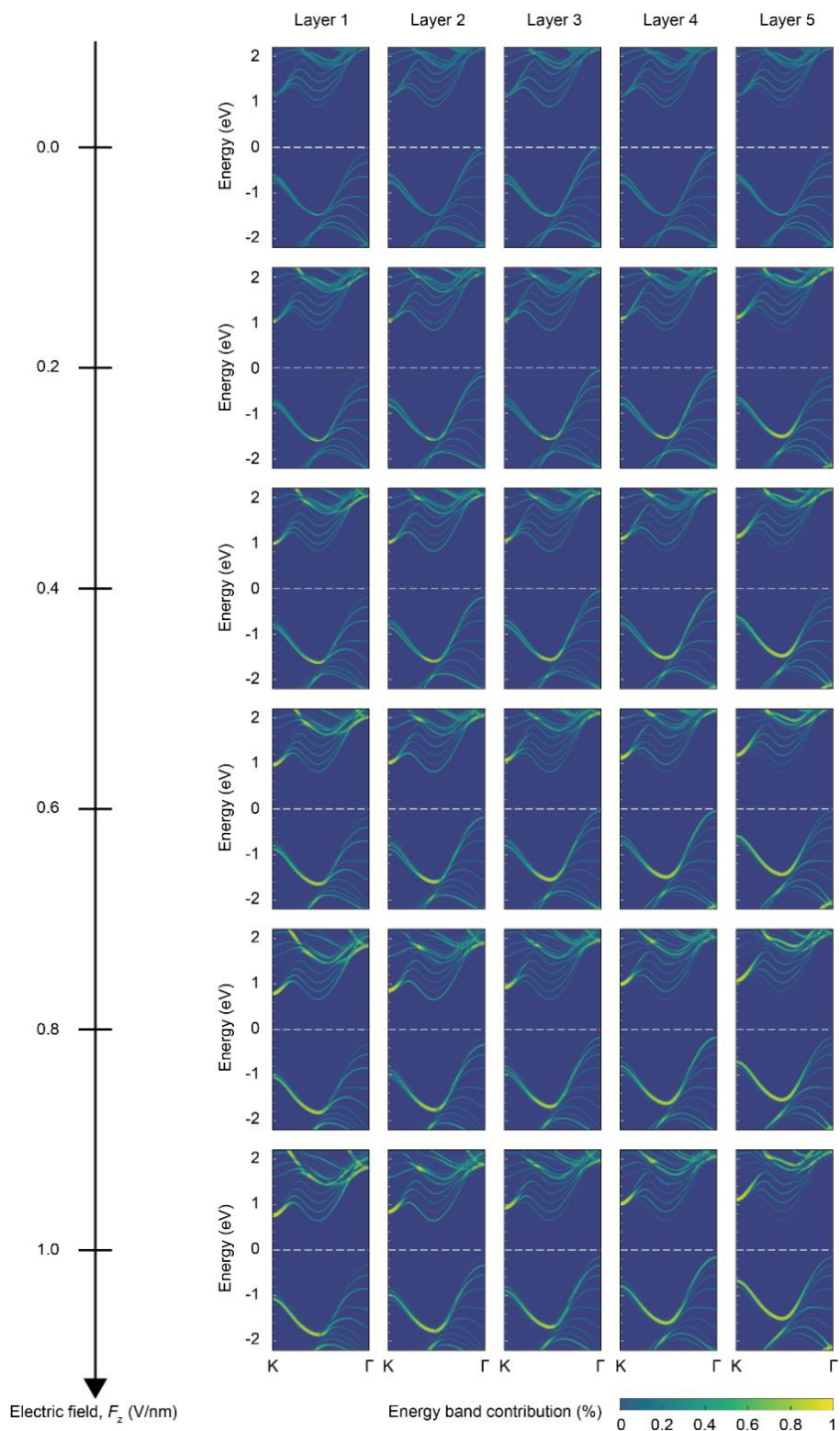

**Figure S3. Layer-decomposed electronic structure evolution in few-layer MoS$_2$ in an external electric field.** As the electric field increases, the layer-decomposed bands begin to split and form the Stark ladders.

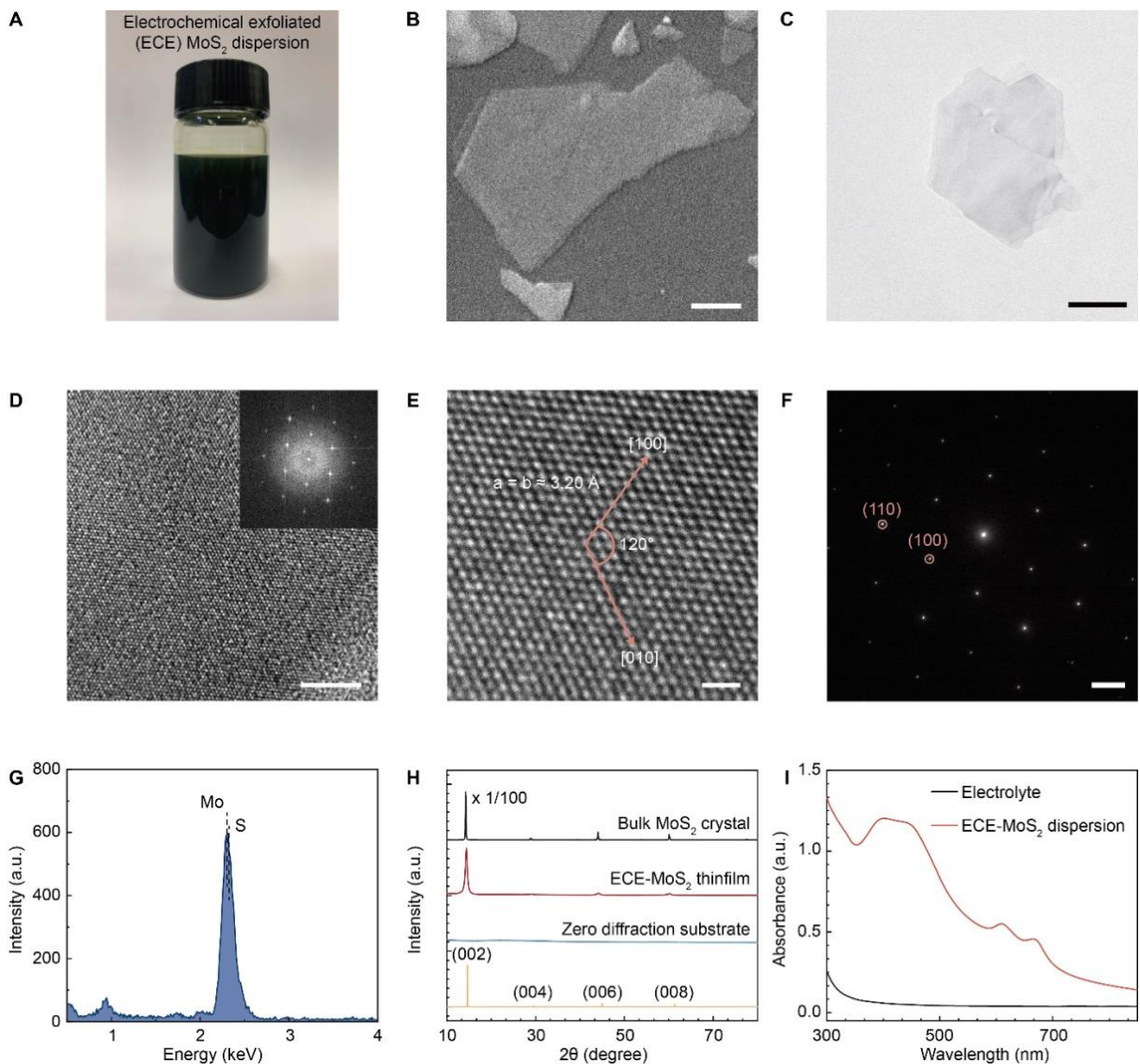

**Figure S4. Characterizations of the solution-processed MoS₂ prepared by electrochemical exfoliation.** (A) Photograph of an electrochemical exfoliated MoS₂ dispersion. (B) Scanning electron microscopic (SEM) image of a typical MoS₂ nanosheet. (C) Transmission electron microscopic (TEM) image, (D) high-resolution TEM image, and (E) the zoomed-in high-resolution TEM image of a typical MoS₂ nanosheet. The inset in (D) is the fast Fourier transform (FFT) pattern of the selected area. The lattice parameters measured in (E) of the MoS₂ nanosheet are about 3.20 Å. (F) Selected area electron diffraction image of the MoS₂ nanosheet, matching the FFT pattern shown in (D). (G) Energy dispersive X-ray characterization of the solution-processed MoS₂, showing strong Mo and S signals. (H) X-ray diffraction characterizations of the solution-processed MoS₂ and the bulk crystal. The expansion of the (002) peak and the weakening of the (004), (006), and (008) peaks indicate an effective exfoliation. (I) Optical absorption spectrum of the electrochemical exfoliated MoS₂ dispersion, showing the characteristic peaks of the 2H semiconducting phase. Scale bars – (B) 200 nm, (C) 200 nm, (D) 5 nm, (E) 1 nm, and (F) 2 nm⁻¹.

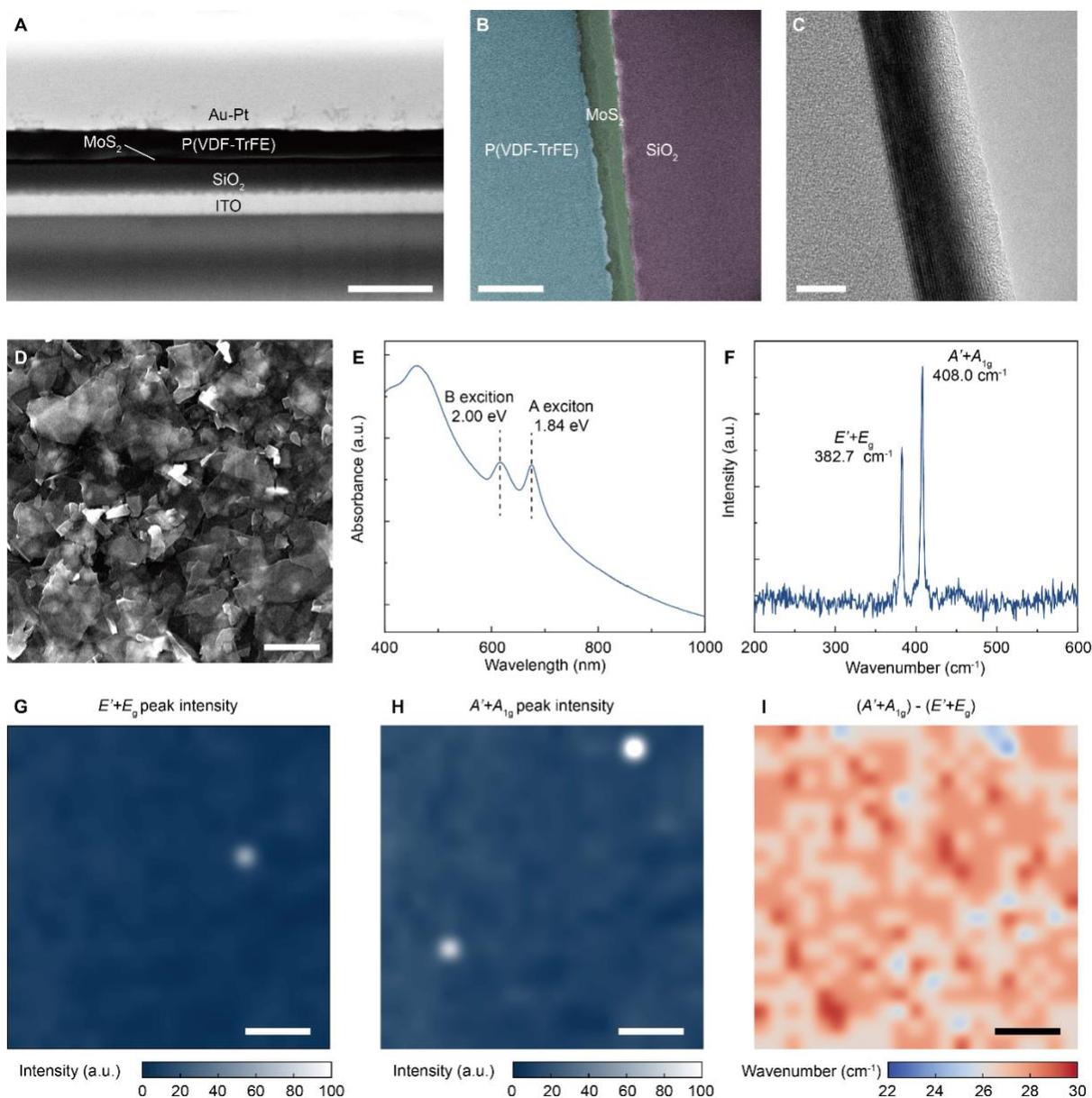

**Figure S5. The ITO/SiO$_2$/MoS$_2$/P(VDF-TrFE)/Au device based on the solution-processed MoS$_2$.** (A) Large-scale cross-sectional scanning transmission electron microscopic (STEM) image of the device, showing 5-layer structure of the device with distinctly clear interfaces. (B) Cross-sectional STEM image of the device, and (C) the zoomed-in cross-sectional STEM image. The stacked atomic nanosheets in the solution-processed MoS$_2$ layer are observed. (D) Top-view SEM image of a deposited solution-processed MoS$_2$ thin-film. The solution-processed MoS$_2$ ensemble consists of randomly stacked MoS$_2$ nanosheets. (E) The optical absorption spectrum of the solution-processed MoS$_2$ thin-film. (F) Raman scattering characterization of the solution-processed MoS$_2$ thin-film. (G-I) Raman scattering mapping of a randomly selected area in the solution-processed MoS$_2$ thin-film – the Raman peak intensity mapping of (G) the $E' + E_\mathrm{g}$ and (H) $A' + A_{1\mathrm{g}}$ vibrational modes, and their gap. The averaged gap is about 26 cm$^{-1}$. Scale bars – (A) 1 µm, (B) 100 nm, (C) 10 nm, (D) 500 nm, and (G)-(I) 20 µm.

**Supplementary References**

[1] Kresse, G. and Furthmuller, J. Efficient iterative schemes for *ab initio* total-energy calculations using a plane-wave basis set. *Phys. Rev. B* **54**, 11169 (1996).

[2] Blochl, P. E. Projector augmented-wave method. *Phys. Rev. B* **50**, 17953 (1994).

[3] Perdew, J. P., Burke, K. and Ernzerhof, M. Generalized gradient approximation made simple. *Phys. Rev. Lett.* **77**, 3865 (1996).

[4] Grimme, S., et al. A consistent and accurate ab initio parametrization of density functional dispersion correction (DFT-D) for the 94 elements H-Pu. *J. Chem. Phys.* **132**, 154104 (2010).

[5] Grimme, S., Ehrlich, S. and Goerigk, L. Effect of the damping function in dispersion corrected density functional theory. *J. Comput. Chem.* **32**, 1456-1465 (2011).

[6] Momma, K. and Izumi, F. VESTA 3 for three-dimensional visualization of crystal, volumetric and morphology data. *J. Appl. Crystallogr.* **44**, 1272-1276 (2011).

[7] Lin, Z. Y., et al. Solution-processable 2D semiconductors for high-performance large-area electronics. *Nature* **562**, 254-258 (2018).

[8] Hamaguchi, C., *Basic semiconductor physics (3rd edition)*. (Springer, 2017).

[9] Hader, J., Linder, N. and Dohler, G. H. **k·p** theory of the Franz-Keldysh effect. *Phys. Rev. B* **55**, 6960 (1997).

[10] Miller, D. A. B., Chemla, D. S. and Schmittrink, S. Relation between Electroabsorption in bulk semiconductors and in quantum wells: The quantum-confined Franz-Keldysh effect. *Phys. Rev. B* **33**, 6976 (1986).

[11] Shi, H. Y., et al. Exciton dynamics in suspended monolayer and few-layer $MoS_2$ 2D crystals. *ACS Nano* **7**, 1072-1080 (2013).

[12] Wang, R., et al. Ultrafast and spatially resolved studies of charge carriers in atomically thin molybdenum disulfide. *Phys. Rev. B* **86**, 045406 (2012).

[13] Wang, G., et al. Excitons in atomically thin transition metal dichalcogenides. *Rev. Mod. Phys.* **90**, 021001 (2018).

[14] Horikoshi, Y., Fischer, A. and Ploog, K. Photoluminescence quenching in reverse-biased $Al_xGa_{1-x}As$/GaAs quantum-well heterostructures due to carrier tunneling. *Phys. Rev. B* **31**, 7859 (1985).

[15] Mendez, E. E., et al. Effect of an electric-field on the luminescence of GaAs quantum wells. *Phys. Rev. B* **26**, 7101 (1982).